\documentclass[10pt,aps,prl,twocolumn,amsmath,amssymb,superscriptaddress]{revtex4-1}
\usepackage[utf8]{inputenc}
\usepackage[T1]{fontenc}
\usepackage{microtype}
\usepackage[dvips]{graphicx}
\usepackage{xcolor}
\usepackage{times}
\usepackage{appendix}
\usepackage{chngcntr}
\usepackage{etoolbox}
\usepackage{lipsum}\usepackage{indentfirst} 
\usepackage{amsmath}
\usepackage{amsfonts}
\usepackage{amssymb}
\usepackage{color}
\usepackage{graphicx}
\usepackage{caption}
\usepackage{subcaption}
\usepackage{mwe}
\graphicspath{{figures/}}
\usepackage[subpreambles=true]{standalone}
\usepackage{import}

\usepackage[normalem]{ulem}
\usepackage{color}

\newcommand{\meanv}[1]{\left\langle #1 \right\rangle}
\newcommand{\bb}[1]{\left( #1 \right)}

\newcommand{\be}{\begin{equation}}
\newcommand{\ee}{\end{equation}}
\newcommand{\bea}{\begin{eqnarray}}
\newcommand{\eea}{\end{eqnarray}}

\newcommand{\mrm}[1]{\mathrm{#1}}

\newcommand{\dd}{\mrm {dd}}
\newcommand{\sr}{\mrm {sr}}

\newcommand{\1}{\uparrow}

\usepackage[T1]{fontenc} 

\begin{document}

\title{Strongly correlated quantum droplets in quasi-1D dipolar Bose gas}
\author{Rafa{\l}~ O{\l}dziejewski}
\author{Wojciech~G{\'{o}}recki}
\author{Krzysztof~ Paw{\l}owski}
\author{Kazimierz~Rz\k{a}\.{z}ewski}

\affiliation{Center for Theoretical Physics, Polish Academy of Sciences, Al. Lotnik\'{o}w 32/46, 02-668 Warsaw, Poland}


\begin{abstract}
We exploit a few- to many-body approach to study strongly interacting dipolar bosons in the quasi-one-dimensional system. The dipoles attract each
other while the short range interactions are repulsive. Solving numerically the multi-atom Schr\"{o}dinger equation, we discover that such systems can exhibit not only the well known bright soliton solutions but also novel quantum droplets for a strongly coupled case. For larger systems, basing on microscopic properties of the found few-body solution, we propose a new equation for a density amplitude of atoms. It accounts for fermionization for strongly repelling bosons by incorporating the Lieb-Liniger energy in a local density approximation and approaches the standard Gross-Pitaevskii equation (GPE) in the weakly interacting limit. Not only does such a framework provide an alternative mechanism of the droplet stability, but it also introduces means to further analyze this previously unexplored quantum phase. In the limiting strong repulsion case, yet another simple multi-atom model is proposed. We stress that the celebrated Lee-Huang-Yang term in the GPE is not applicable in this case.
\end{abstract}


\maketitle
The advent of degenerate quantum gases consisting of atoms interacting via strong long-range dipolar forces has brought a fascinating perspective of the study of new quantum states of matter. During the last three years, self-bound droplets have been unexpectedly observed~\cite{Schmitt2016}, so was a long-awaited roton excitation~\cite{chomaz2018} leading to the subsequent realization of dipolar supersolids~\cite{Modugno2019,Pfau2019,Chomaz2019}.

Quantum droplets in three dimensions in ultracold dipolar gases and Bose-Bose mixtures are stable due to the many-body corrections to the ground state energy that introduce an additional repulsion in the system given by a seminal Lee-Huang-Yang (LHY) term~\cite{Petrov2015}. Usually negligible, it determines the properties of the system when repulsive and attractive mean-field contributions almost cancel out each other. Since the first observation of quantum droplets~\cite{Kadau2016}, intense theoretical~\cite{Wachtler2016,Baillie2016,saito2016path,OldziejewskiJachymski2016,baillie2017collective} and experimental~\cite{Barbut2016,Schmitt2016,Chomaz2016,Barbut2018,bottcher2019quantum,Tarruell2018,Tarruell2018a,Fattori2018}
efforts have been devoted to characterizing their properties in both dipolar gases and Bose-Bose mixtures. 

Quantum liquids are also believed to be present in lower dimensions.
For Bose-Bose mixtures, beyond mean-field effects are enhanced due to geometrical confinement, and as a consequence, the liquid state becomes even more ubiquitous and remarkable~\cite{Petrov2016,Malomed2018}. In this case, LHY contribution causes the additional attraction present in the system.
Unfortunately, calculations of similar corrections in the dipolar gases proved to be more demanding.
For a quasi-two-dimensional system, the result is non-universal and depends on features of the confinement~\cite{OldziejewskiJachymski2018}.
For a quasi-one-dimensional dipolar gas, the attractive beyond mean-field contribution was calculated, supporting the existence of dipolar quantum droplets in such a geometry~\cite{Santos2017}.
However, in this case, the correction comes from the possible transverse excitations of the system, which strongly suggests that this result is also non-universal.

In this Letter we present a quasi-1D quantum droplet of a different kind. Our new self-bound state emerges in the system of $N$ polarized cold dipoles with the net repulsive interaction. We focus on a regime which is far beyond the Bogoliubov approximation, and therefore in which LHY correction is not applicable. We start with a few-body system, sufficiently small to perform the numerical diagonalization. We show, that due to the interplay between short range van der Waals interaction and non-local dipolar forces, the ground state of the system is either a dipolar bright soliton (net attractive interaction) or another self-bound state, called here the quantum droplet (net repulsive interaction).

Next, we move to larger systems. As exact numerics is impossible for thousands of atoms, we change completely our framework. Referring to the energy functional discussed by Lieb~\cite{LiebLiniger1963} and using the local-density approximation, we propose a new equation for a density amplitude. This equation is called here the Lieb-Liniger Gross-Pitaevskii equation (LLGPE). Although our approach does not assume a universal orbital for particles, it reduces to the GPE in the limit of weak interactions or large densities. In another extreme, of the infinitely strong repulsive interaction, it was studied in~\cite{Kolomeisky2000,Baizakov2009}. Using our equation, we calculate the quantum phase diagram and show the characteristic flat-top shape of the density of the gas in the quantum droplet state. Finally, we use a many-body variational Ansatz to show that in the infinitely strong repulsive interaction limit the quantum droplet can be understood as the self-confined Tonks-Girardeau gas.

\begin{figure}[h!]
\begin{centering}
		\includegraphics[width=0.42\textwidth]{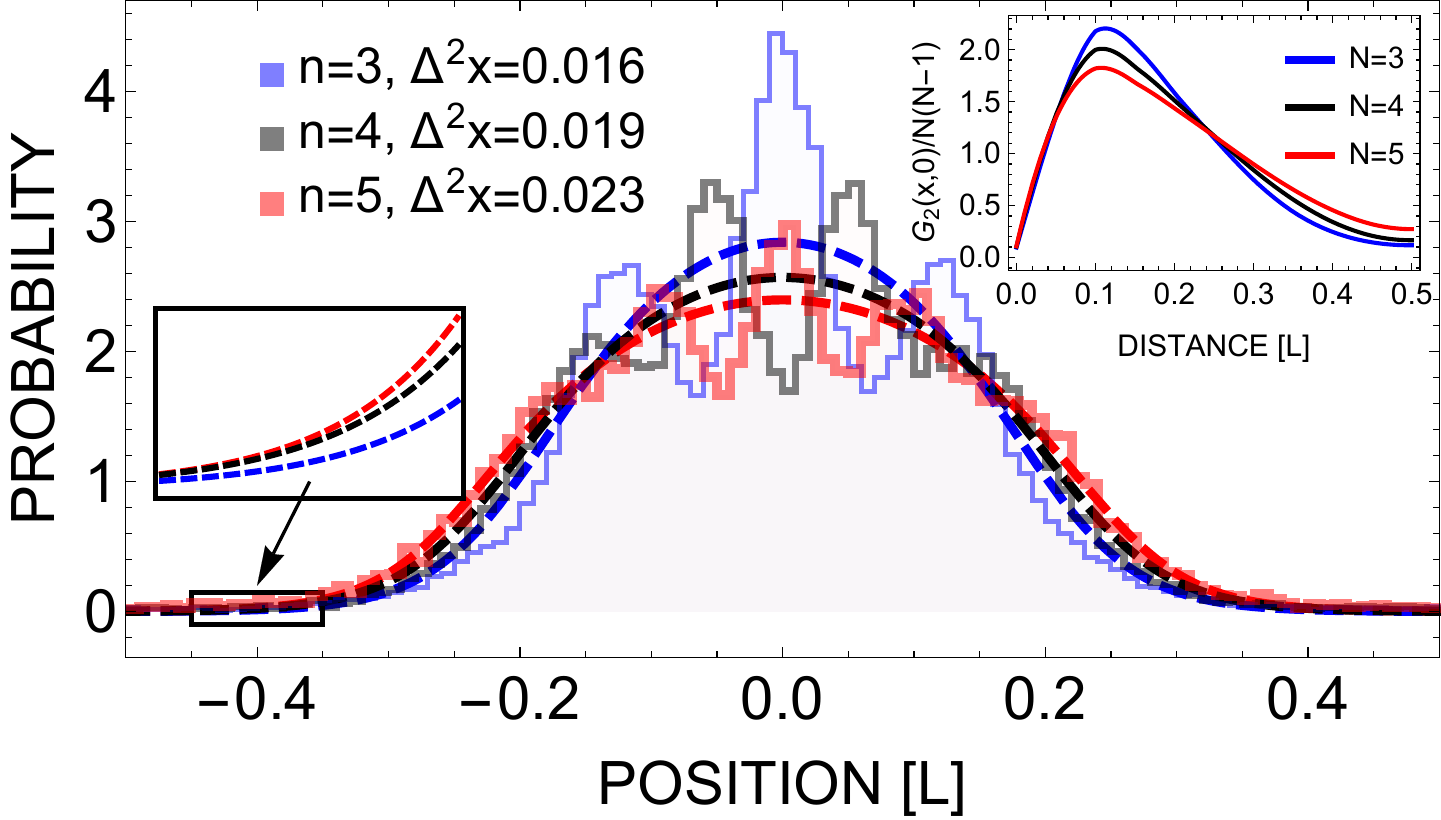}
		
		\includegraphics[width=0.42\textwidth]{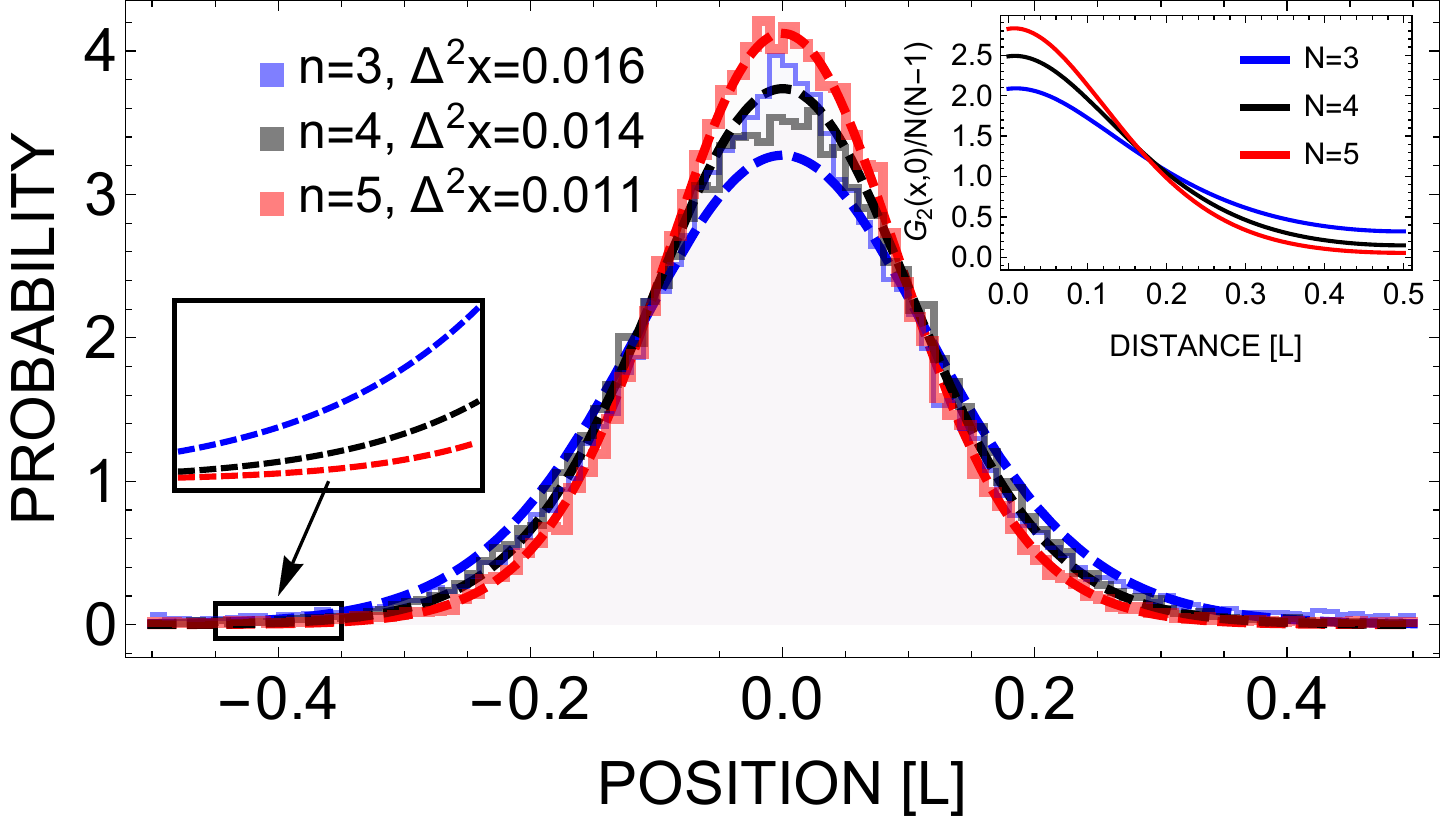}

\par\end{centering}
\caption{(color online) Probability density histograms of particles' positions for the ground state for $N=$ 3 (blue solid), 4 (black solid) and 5 (red solid) atoms with $f_{\mrm {dd}}=0.9$, $g_\dd=135$ and $\sigma=0.2$ (top) and for $f_{\mrm {dd}}=20$, $g_\dd=15$ and $\sigma=0.2$ (bottom) compared to solutions of Eq. \eqref{eqddllgpe} with the same parameters (dashed lines, same color coding). The particles' positions were drawn with the Metropolis algorithm~\cite{Oldziejewski2018a}. 
Note that the density value in the middle depends on the parity of number of particles.
\label{histsoldrop}}
\end{figure}

We consider $N$ dipolar bosons confined in both transverse directions $\hat{y}$ and $\hat{z}$ 
with a tight harmonic trap of a frequency $\omega_\perp$. 
Multi-particle wave-function is approximately Gaussian in tight directions for all variables.
In the longitudinal direction $\hat{x}$ the space is assumed to be finite, with the length $L$. Atoms are polarized along the $\hat{x}$ axis in head-to-tail configuration~\cite{Lahaye2009}. 
Our quasi-1D system is governed by the Hamiltonian:
\be
\begin{split}
\hat{H}=&-\frac{\hbar^2}{2m}\int dx \, \hat{\psi}^\dagger(x)\nabla^2\hat{\psi}(x) \\ & +\frac{1}{2}\int dxdx'\, \hat{\psi}^\dagger(x)\hat{\psi}^\dagger(x')V_{\mrm {eff}}(x-x')\hat{\psi}(x)\hat{\psi}(x')\label{hamgen}
\end{split}
\ee
with $\hat{\psi}(x)$ being a bosonic field operator.
The effective potential consists of the long-range dipolar part and the short-range part, namely ${V_{{\rm eff}}(x) =V_{{\rm dd}}(x)+V_{{\rm sr}}(x)}$. The quasi-1D dipolar potential reads $V_\dd(x)=-\frac{\mu_0 D^2}{2\pi l_{\perp}^2} \frac{v_\dd\left(x/l_{\perp}\right)}{l_{\perp}}$, where ${v_\dd(u)=\frac{1}{4}\left(-2|u|+\sqrt{2\pi}(1+u^2)e^{u^2/2}{\rm{Erfc}}\left(|u|/\sqrt{2}\right) \right)}$ is normalized by $\int v_\dd(u)du=1$ and $l_{\perp}=\sqrt{\hbar/m\omega_{\perp}}$. Here, $D$ is a value of atomic dipole moment and $\mu_0$ is the vacuum permeability. This effective quasi-1D potential comes from integration of the full 3D dipolar interaction over both transverse variables~\cite{deuretzbacher2010ground}. The singular part coming from this integration~\cite{deuretzbacher2013erratum} is incorporated within the short range interaction. 

In this Letter, the atoms repel each other on the short distance. Thus, we use the usual model of short-range interactions $V_{{\mrm {sr}}}(x)=\frac{\hbar^2 a}{m l_{\perp}^2} \delta(x)$
with $a\geq 0$ mimicking the scattering length, which can by tuned in experiments by Feschbach resonances. Below we use box units where $L$, $\hbar/L$ and $\hbar^2/mL^2$ are the units of length, momentum and energy respectively. In addition, we also define coefficients $g=\frac{\hbar^2 a}{m l_{\perp}^2}$, $g_\dd=\frac{\mu_0 D^2}{2\pi l_{\perp}^2}$, the aspect ratio $\sigma=l_{\perp}/L$ and the rescaled function $v^\sigma_\dd(x):=\frac{1}{\sigma}v_\dd(x/\sigma)$, so that finally the effective potential takes a compact form ${V_{\mrm {eff}}(x)=-g_{\dd}v^{\sigma}_{\dd}(x)+g\delta(x)}$.

We are interested in how the properties of the ground state of the system depend on the strength of interactions. We manipulate two of the parameters of interactions: the ratio between dipolar and contact interactions $f_{\mrm{dd}}=g_\dd/g$ and $g_{\mrm {dd}}$ itself. We expect that for $f_{\mrm {dd}}>1$ (net attraction) the ground state has negative energy and atoms form a self-bound state similar to the bright solitons studied earlier in ultracold gases.~\cite{Edmonds2017}. On the other hand, the bound ground state for $f_{\mrm {dd}}<1$ (net repulsion) was also presented in quasi-1D systems within the modified mean-field analysis in the extreme case of $f_{\mrm {dd}}=0$ (with $g_\dd>0$ and $g\rightarrow \infty$)~\cite{Baizakov2009}. With our methods, we can investigate features of the system in a many-body manner across the whole range of parameters.
Chiefly, we are interested in the potentially transitional change of the ground state while crossing $f_{\mrm {dd}} \sim 1$, keeping $g_\dd$ constant.
 
 We attempt to find a few-body ground state for both $f_{\mrm{dd}}<1$ and $f_{\mrm{dd}}>1$ with negative energy and characteristic width smaller than $L$, so that we can study the spatial properties of such a bound-state. We access the many-body eigenstates of the Hamiltonian by diagonalization using the Lanczos algorithm~\cite{lanczos1950iteration}. We assume the periodic boundary conditions for a numerical convenience~\cite{oldziejewski2018many,Oldziejewski2018a}. The test of convergence of our numerical method is discussed in the Supplemental Material.

As an example, in Fig. \ref{histsoldrop}, we present probability density histograms for the net repulsive interactions (top panel, $f_{\mrm {dd}}=0.9$)  and the net attractive interactions (bottom panel, $f_{\mrm {dd}}=20$). We select the interaction parameters so that the histograms have a similar width for attractive and repulsive scenarios. Both histograms are obtained by drawing particles' positions from the many-body probability distribution with the Metropolis algorithm and aligned by rotating them such that their center of mass point in the same direction \cite{Oldziejewski2018a}. We observe two spatially localized bound-states with completely different properties. Firstly, for $f_{\mrm{dd}}=0.9$, the width increases as the number of atoms grows, which is the opposite for $f_{\mrm{dd}}=20$. In the first case  
we observe local peaks whose number agrees with the number of particles, whereas in the latter only a single central peak is observed. 
The above features strongly resemble the quantum droplets and bright solitons differences discussed in the recent papers about dipolar systems and Bose-Bose mixture~\cite{Santos2017,Malomed2018}. Therefore, we name the first case as a droplet-like and the second one as a soliton-like solution.  

The origin of the differences between the two quantum phases is evident when we look at quantum bunching properties, characterized by the normally ordered second-order correlation function $G_2(x,x') := \langle \Psi^{\dagger}(x)\Psi^{\dagger}(x')\Psi^{}(x')\Psi^{}(x) \rangle$. This function is crucial for the average energy of any system described by Eq. \eqref{hamgen}: 
\be
\begin{split}
\meanv{\hat{H}}=&-\frac{1}{2}\int dx\,\langle \hat{\psi}^\dagger(x)\nabla^2\hat{\psi}(x)\rangle\\
&+\frac{1}{2}\int dxdx'\,\,G_2(x,x')V_{\mrm {eff}}(x-x')
\end{split}\label{eqaverage}
\ee
In particular, we see that in order to develop a bound-state for $f_{\mrm {dd}}<1$, the second-order correlation function needs exhibiting a local minimum for atoms at the same place, to decrease the contribution from the short-range repulsion which requires sufficiently strong interactions.

With our diagonalization technique, we have an access to the function $G_2(x,x')$ shown in the insets of Fig. \ref{histsoldrop}. A dramatic difference between both situations can be found: atoms avoid each other in quantum droplets ($\frac{G_2(0,0)}{N(N-1)}\ll 1$) and tend to bunch in solitons ($\frac{G_2(0,0)}{N(N-1)} >1$)~\footnote{Note, that in our example the system is translationally invariant. Somewhat unusual normalization factor comes from the fact that we compare $G_2$ function with the density probability given by the histograms.}).

To study larger systems around $f_{\mrm {dd}}\sim 1$, we propose a new framework, independent from the diagonalization scheme but resulting from our previous microscopic considerations. As in the case of droplets $G_2(x,x')$ exhibits the anti-bunching, one could not assume a universal orbital describing all particles as one does in the GPE derivation. Instead, the density functional approach similar to the Thomas-Fermi approximation (TFA) for fermions can be constructed in the following way~\cite{SM}.
If only dipolar interactions change much slower than a range of anticorrelations in $G_2(x,x')$, namely that $N \sigma \gg d$ with $d$ being a width (FWHM) of a solution density $\rho(x)=N\left| \psi(x)\right|^2$, one can treat them classically (like a trapping potential in TFA). Accordingly, we assume, that locally the atoms obey the ground state from the Lieb-Liniger model (like the kinetic energy of fermions in TFA). 
 We approximate the Lieb-Liniger energy as a density function in a very simplified way, namely as {$e_{\mrm {LL}}=\frac{gN(N-1)}{2}\frac{|\psi|^6}{|\psi|^2+\frac{3g}{N\pi^2}}$}~\cite{SM}.
With that, the energy functional for our system reads:
\be
\begin{split}
E&=  \int dx \left [\frac{N}{2}\left |\nabla \psi \right|^2+\frac{gN(N-1)}{2}\frac{|\psi|^6}{|\psi|^2+\frac{3g}{N\pi^2}} \right ] \\
& - \frac{g_\dd N(N-1)}{2}\int dx dx'\left |\psi(x) \right|^2 v_\dd^\sigma(x-x')\left |\psi(x') \right|^2
\end{split}
\ee
where $\int dx \left| \psi(x)\right|^2 =1$. Then, using variational calculus, we arrive at a new equation for $\psi (x)$:

\begin{equation}\label{eqddllgpe}
\begin{split}
\mu &\psi(x) =- \frac{N}{2}\frac{\partial^2 }{\partial\,x^2}\psi(x)+f_{\mrm{LL}}\big[ \psi(x)\big]
 \\ & -g_\dd N(N-1)\int dx' v_\dd^\sigma(x-x')\left| \psi(x')\right|^2 \psi(x),
\end{split}
\end{equation}
where $f_{\mrm{LL}}\big[ \psi(x)\big]=\frac{\delta e_{\mrm {LL}}}{\delta \psi^{*}}$ and $\mu$ is a Lagrange multiplayer.

Note that the above equation reduces to the two equations previously studied in the literature. In the limit of very weak contact interactions $g\rightarrow 0$ we restore the standard GPE, while for $g\rightarrow \infty$ (the Tonks-Girardeau limit) we obtain the equation investigated earlier in~\cite{Kolomeisky2000,Baizakov2009}. Therefore, we name Eq. \eqref{eqddllgpe} the Lieb-Liniger GPE (LLGPE). However, note that our reasoning does not originate from the usual product state approximation for the many-body wave function for ultracold bosons and does describe the density amplitude of $N$ particles rather than a universal orbital occupied by all particles. We emphasize that we use a simplified energy density functional for the ground state, which is an approximation of the full Lieb-Liniger expression, see for instance~\cite{Miguzzi2017,Miguzzi2018}.
\begin{figure}[h!]
\begin{centering}
		\includegraphics[width=0.43\textwidth]{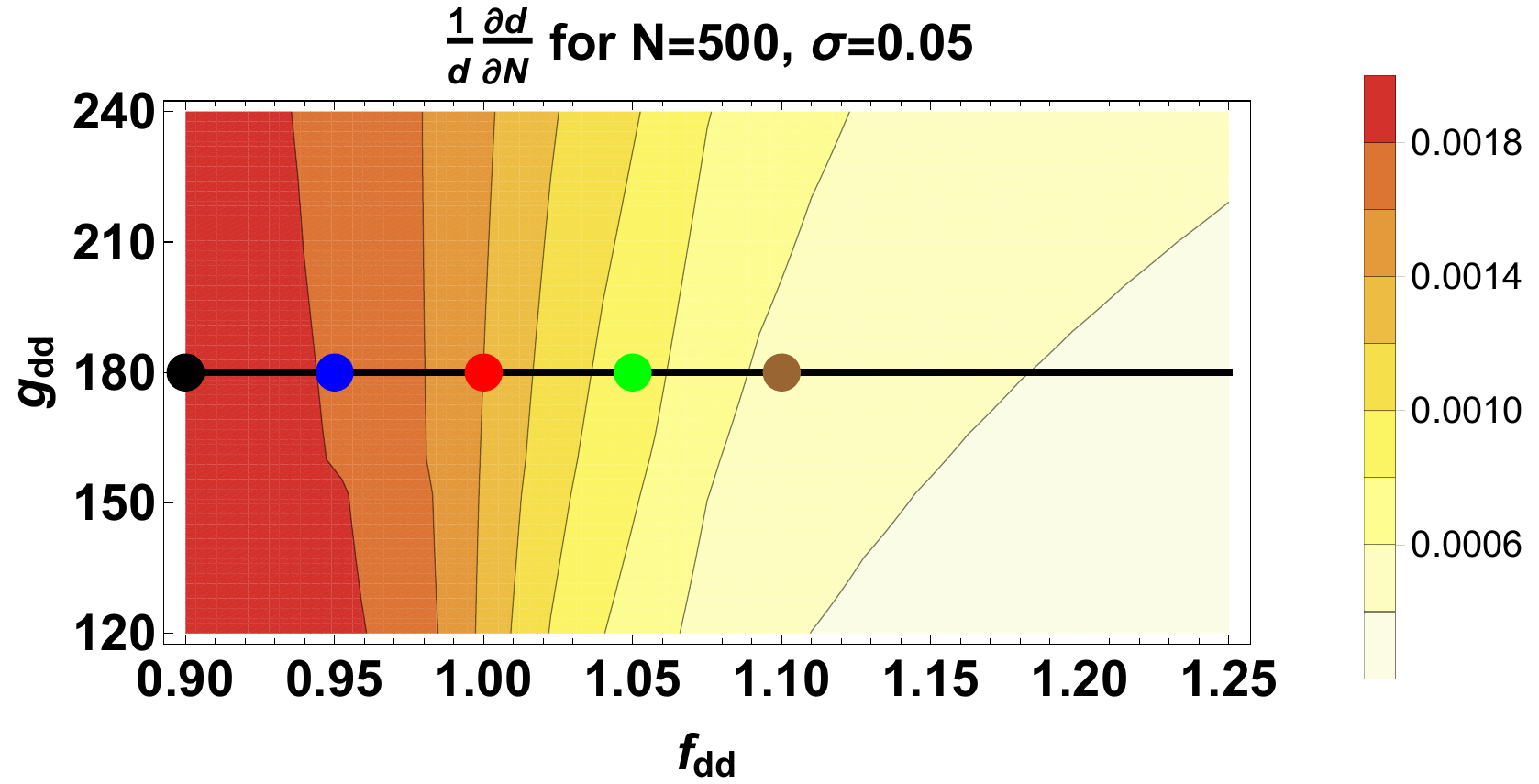}
		
		\includegraphics[width=0.43\textwidth]{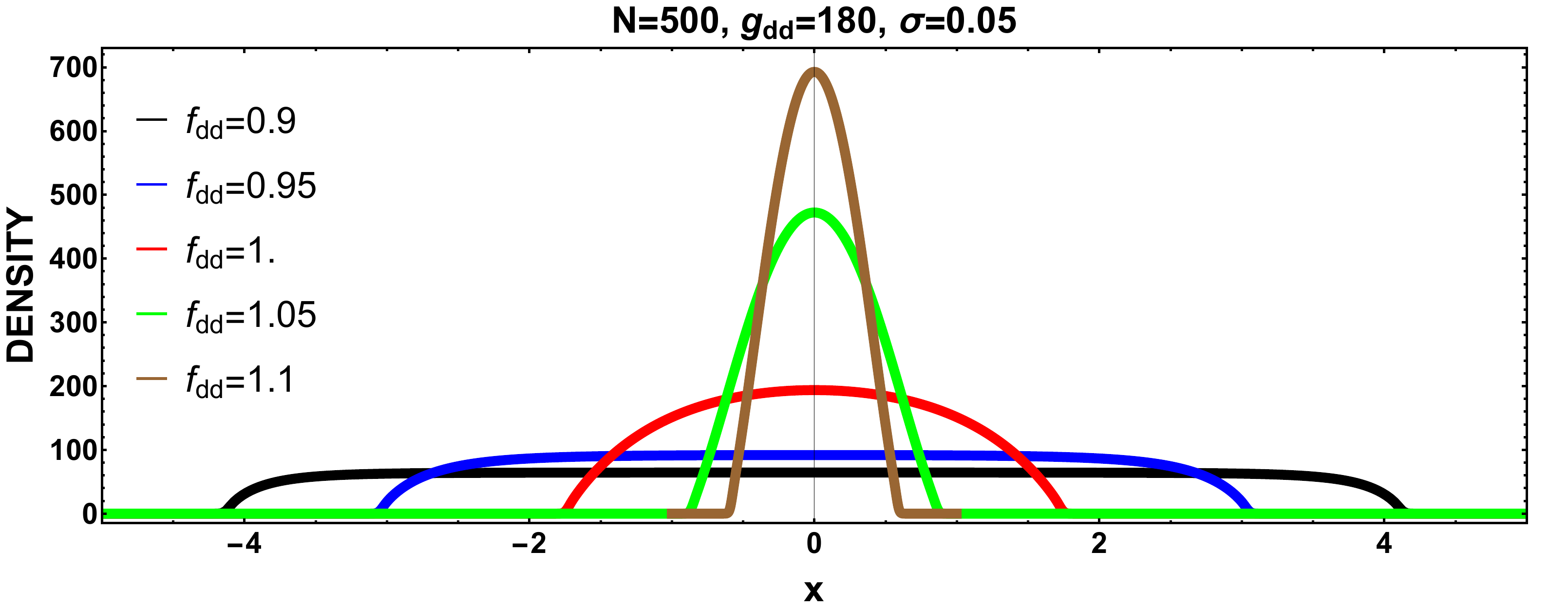}
\par\end{centering}
\caption{(color online) Top: First derivative of a width $d$ of the solution of Eq. \eqref{eqddllgpe} over $N$ as a function of the parameters $f_{\mrm {dd}}$ and $g_\dd$. Bottom: Density for different $f_{\mrm {dd}}$ with parameters as for the colour points from the top panel}\label{diagram}
\end{figure}

We aim to solve the Eq. \eqref{eqddllgpe} for exactly the same parameters as for the few-body case. For this, we use the imaginary time evolution (ITE) technique. 
In Fig. \ref{histsoldrop}, we compare the outcomes from the ITE with the results from the few-body calculations ($\rho(x)$ from the ITE are marked by dashed lines).
Most importantly, LLGPE captures the same $N$ dependence as the few-body considerations. We see that especially for N=5 both approaches correspond to each other in a satisfying way both for soliton-like states and droplet-like states. In the first case, we also confirm sech-shape of the solution~\cite{Edmonds2017}, so we will call it hereafter a bright dipolar soliton. Remembering that Eq. \eqref{eqddllgpe} should work for a large number of atoms where $N \sigma \gg d$ rather than $N \sigma > d$ as in the few-body case, this is an excellent agreement between two different methods. It seems that the LLGPE describing macroscopic density amplitude of the many-body system reproduces few-body behaviour of the system well enough. The agreement between these two approaches gives yet another example where already not very many particles mimic the many-body limit as for instance in~\cite{wenz2013few,grining2015crossover}. Our comparison provides good reasons to focus on the features of LLGPE solutions themselves.

As we discussed before for the small system analysis, droplet-like states get wider as $N$ grows, but bright solitons shrink. Then, it is instructive to consider the first derivative of a width $d$ of the solution of Eq. (\ref{eqddllgpe}) over $N$ as a function of the parameters $f_{\mrm {dd}}$ and $g_{\dd}$. In Fig. \ref{diagram} we present such an analysis. As we have expected, for $f_{\mrm {dd}}<1$ the derivative is positive. Then, it decreases abruptly for $f_\dd>1$~\footnote{Note that there is no sign flip of the derivative as one would expect for a bright-solitons for a purely attractive system. That would be the case for $f_\dd\rightarrow\infty$ (so that $g\rightarrow 0$).}. 
The rapid drop of the derivative of the width is associated with a change of a density profile $\rho(x)=N\left| \psi(x)\right|^2$ as one crosses $f_{\mrm {dd}}\approx 1$. We observe a transformation of the density from a flat-top to sech-shape profile. 

\begin{figure}[]
\begin{centering}
		\includegraphics[width=0.46\textwidth]{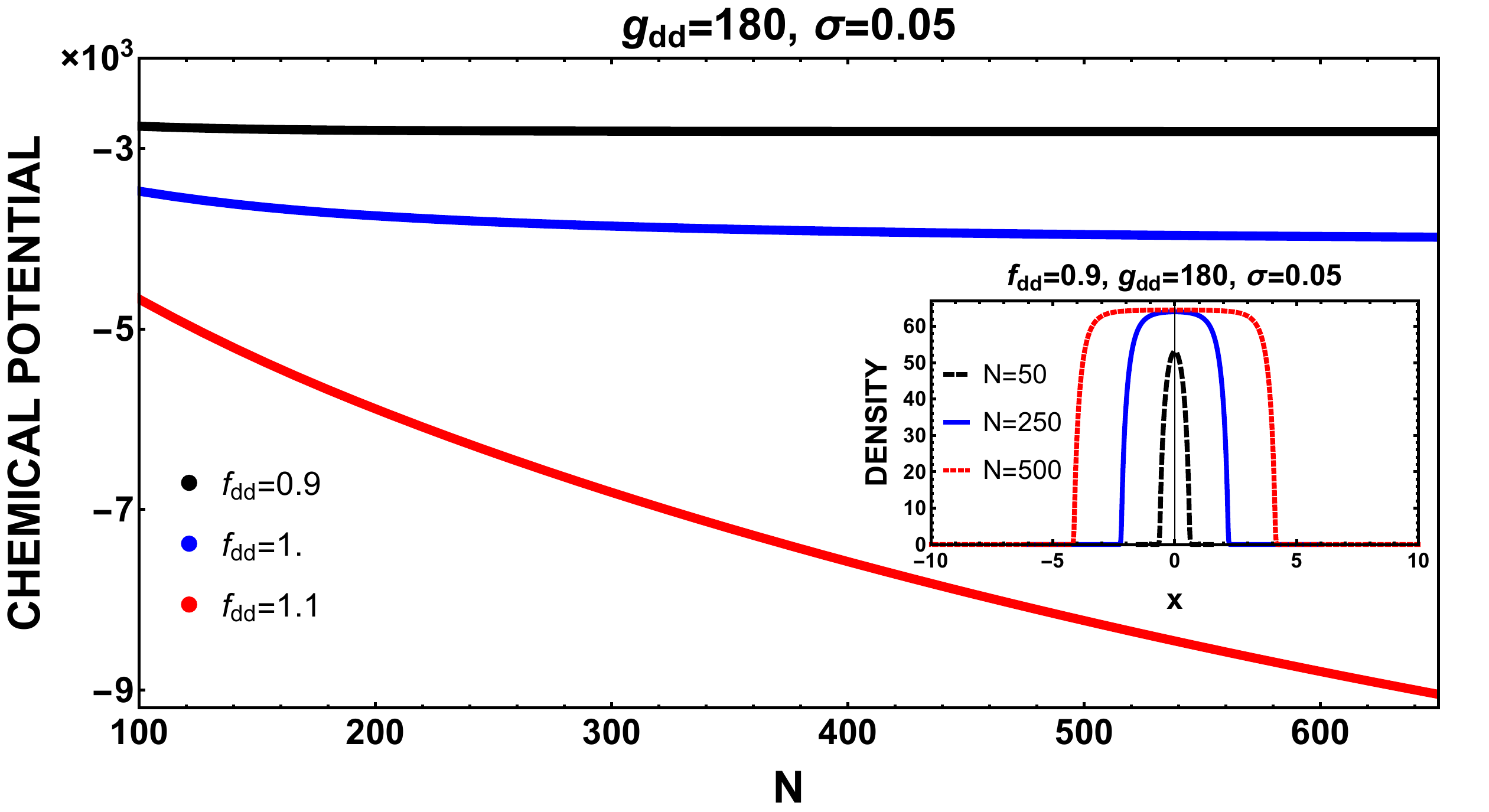}
\par\end{centering}
\caption{(color online) Top: Chemical potential $\mu$ as a function of particles number $N$ for the parameters as for the colored points in Fig. \ref{diagram}. Inset: Density plot for different $N$ and parameters as for the black point in Fig. \ref{diagram}.}\label{chemic}
\end{figure}

Then, we examine the dependence of the solution on the particle number $N$. In the top panel of Fig. \ref{chemic}, we show the chemical potential $\mu=\frac{\partial E}{\partial N}$ as a function of $N$ for the same coupling strengths as in Fig. \ref{diagram}. For all cases, the total energy of the system and the chemical potential are negative, entailing that the states are self-bound. For small $N$ the chemical potential decreases for all $f_{\mrm {dd}}$, for higher $N$ it becomes constant for $f_{\mrm {dd}}\leq 1$ and linear for $f_{\mrm {dd}}> 1$. The constant chemical potential is a hallmark of a droplet. We observe also that as $N$ increases, the density $\rho(x)$ of droplets changes to a flat-top shape (inset in Fig. \ref{chemic}). Note, that even though there is no critical number of particles below which a state is not stable in the empty 1D space, one has to remember that for a very small $N$ it may happen that the solution of LLGPE does not satisfy $N \sigma \gg d$, and thus existence of the corresponding physical state of the system is not guaranteed. 

\begin{figure}[h!]
\begin{centering}
		\includegraphics[width=0.4\textwidth]{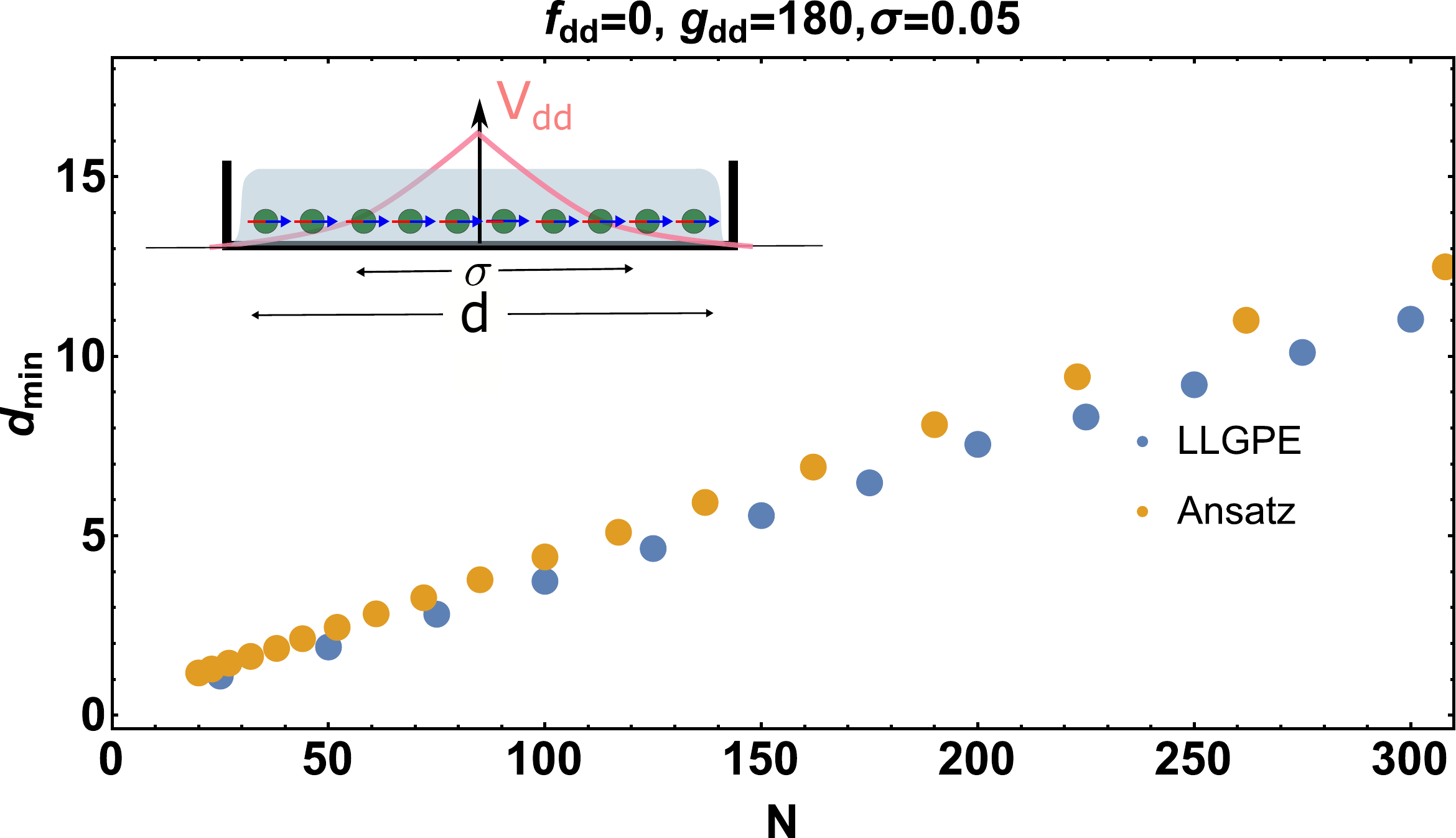}
\par\end{centering}
\caption{(color online) Droplet width $d_{\mrm {min}}$ as a function of $N$ obtained from the numerically exact solution of Eq. \ref{eqddllgpe} (blue points) or the many-body Ansatz (yellow points) and for $f_\dd \rightarrow 0$, $\sigma=0.05$ and $g_\dd=180$. Inset: Artist's view of the droplet in Tonks-Girardeau regime.}\label{ansatz}
\end{figure}

Previously, we have deduced from microscopic properties of few-body systems the density amplitude equation describing self-bound quantum droplets. This equation assumes locally the Lieb-Liniger ground state. For $g\rightarrow \infty$, the ground state becomes the Tonks-Girardeau solution. Therefore, we propose a following many-body Ansatz for the droplet state for $g\rightarrow \infty$. We assume that the droplet state is a ground state of $N$ infinitely repulsive bosons in a hard-wall box of length $d$ that is a variational parameter~\cite{SM}. We compare our numerical minimalization findings with a solution of Eq. (\ref{eqddllgpe}) obtained with ITE for $g \rightarrow \infty$, $\sigma=0.05$ and $g_\dd=180$ in Fig. \ref{ansatz}. We see a good agreement between both methods for the width of a droplet. Our Ansatz reflects the structure of the droplet --- the atoms stay in a compact group thanks to the dipolar attractive forces, but locally they can strongly repel (see inset of Fig. \ref{ansatz}). The comparison between the many-body Ansatz for droplets in the Tonks-Girardeau limit and corresponding solutions of LLGPE suggests that for $g<\infty$ and $f_\dd<1$ a proper many-body Ansatz for the droplet ground state in 1D can be constructed similarly by using the Lieb-Liniger model.

In conclusion, we have found a novel dipolar quantum droplet in the regime of strongly interacting bosons for the net repulsive two-body interactions. We have also studied the droplet - bright soliton transition when changing the effective interactions in the system from being repulsive to attractive. By considering general properties of the Hamiltonian and microscopic features of the few-body system, we have argued that the existence of the 1D droplet state requires a local minimum of the normally ordered second-order correlation function $G_{2}(x,x')$ for $x=x'$. These requirements lead to a new equation for the density amplitude (LLGPE) that incorporates local correlations of the Lieb-Liniger model. Therefore, quantum droplets found in this Letter are far away from the usual mean-field theory even with seminal first order corrections as their effective description contains the fully quantum solution of the Lieb-Liniger model even for strong interactions. 
The underlying microscopic mechanism of their presence comes from the competition of short-range repulsion and non-local attraction in the system. In contrast to the recent findings with LHY-like term in~\cite{Santos2017}, we have found a new self-bound state for any number of atoms both in the few-body system and in LLGPE. 


Our findings also motivates a similar few-body approach to strongly interacting systems in higher dimensions. It may stimulate the research to look for quantum droplets for strongly interacting systems, like electric dipoles, where the standard beyond mean-field approximations are not valid. 

\begin{acknowledgments}
We thank Piotr Grochowski for his careful and critical reading of the manuscript. This work was supported by the (Polish) National Science Center Grant 2015/19/B/ST2/02820 (RO, WG and KR). Center for Theoretical Physics is a member of KL FAMO.
\end{acknowledgments}

\newpage
\onecolumngrid
\appendix
\section*{Supplemental material for 'Strongly correlated quantum droplets in quasi-1D dipolar Bose gas'}
\section{Derivation of Lieb-Liniger Gross-Pitaevskii equation}
We consider $N$ bosons interacting via potential $V_{{\rm eff}}(x) =V_{{\rm sr}}(x)+V_{{\rm dd}}(x)$, where $V_{{\rm sr}}(x)$ is a short-range repulsive interaction (modeled by $g\delta(x)$) and $V_{{\rm dd}}(x)$ is a long-range attractive interaction (here the dipol-dipol interaction) with characteristic range $\sigma$. The Hamiltonian can be written in general as:
\begin{equation}
\hat{H}=\hat{T}+\hat{V}_\sr+\hat{V}_\dd\label{full}
\end{equation}
with $\hat{T}$ being the kinetic energy. We are interested in the regime for which $V_{{\rm dd}}(x)$ is sufficiently strong, such that the ground state of the above Hamiltonian is the bound state. In this section, we propose and justify the simplified formalism,
based on applying the exact solution of the Lieb-Liniger model to the local-density approximation approach, which allows us to investigate the energy and density profile of this ground state.

First, we need to clarify the center of mass issue. As the Hamiltonian \eqref{full} is translationally invariant, therefore its ground state has no well-localized center of mass and the ground state density has to be completely flat $\langle \hat\psi^\dagger(x)\hat\psi(x)\rangle={\rm const}$. It does not mean that there are no self-bound states though. 
This is the same problem as in the simpler case of $N$ atoms interacting only via short range attractive potential. 
In that case, the ground state is a bright soliton, but, as here, with an ambiguous  center of mass position. 
As our goal is to find and study the self-bound state,  we would like, in line with a typical experiment, break the symmetry.
The typical way of breaking the symmetry in theoretical considerations is to mimic a measurement of few particles (with random positions $\{x_1,...,x_K\}$), after which the bound state may appear in the higher order correlation function $\langle \hat\psi^\dagger(x_K)...\hat\psi^\dagger(x_1)\hat\psi^\dagger(x)\hat\psi(x)\hat\psi(x_1)...\hat\psi(x_K)\rangle$ instead of $\langle \hat\psi^\dagger(x)\hat\psi(x)\rangle$ (see for instance \cite{Javanainen1996}).
However, for the purpose of this derivation, we use another approach (leading to the analogous results). Instead of \eqref{full}, we consider
\begin{equation}
\hat{H}_{\epsilon}=\hat{H}-\hat{T}_{CM}-\hat\epsilon
\end{equation}
where $\hat{T}_{CM}$ is the kinetic energy of the center of mass and $\bb{-\hat\epsilon}$ is an infinitesimally small, well-localized external attractive potential. Therefore, also its ground sate is well-localized and from now we will discuss the features of the ground state of $\hat{H}_{\epsilon}$ and by $\langle\cdot\rangle$ we will understand the expectation value in this state.

Let us name by $d$ the full width at the half maximum of the  bound ground state. As we consider repulsive contact interaction, density distributions of any two particles are anticorrelated at small distances. It corresponds to the hole in $G_2(x,0)$ function around $x=0$ (see FIG. 1 in main text), whose typical width is of the order of $d/N$. For much bigger distances we have:
\begin{equation}
\label{corel}
\forall_{|x-y|\gg d/N}\,\langle \hat\psi^\dagger(x)\hat\psi^\dagger(y)\hat\psi(x)\hat\psi(y)\rangle\approx\langle\hat\psi^\dagger(x)\hat\psi(x)\rangle\langle\hat\psi^\dagger(y)\hat\psi(y)\rangle
\end{equation}
Now, if the typical distance between neighboring particles is much smaller than the characteristic range of the non-local potential $V_{\rm dd}$, namely \textbf{if $d$ satisfies $d/N\ll \sigma$}, then one may find an intermediate length scale $l$, i.e. $d/N\ll l\ll \sigma$, such that:
\begin{equation}
\begin{split}
\label{assume}
\bullet\quad&\forall_{x'\in[x,x+l]}\, V_\dd(x')\approx V_\dd(x)\\
\bullet\quad&\forall_{x'\in[x,x+l]}\,\langle \hat\psi^\dagger(x')\hat\psi(x')\rangle \approx \langle\hat\psi^\dagger(x)\hat\psi(x)\rangle\\
\bullet\quad&\int_{x}^{x+l}dx'\,\langle\hat\psi^\dagger(x')\hat\psi(x')\rangle\gg 1
\end{split}
\end{equation}
Firstly, this implies that dipolar interactions may be treated classically:
\begin{equation}
\frac{1}{2}\int dxdy\,\langle\hat\psi^\dagger(x)\hat\psi^\dagger(y)V_{\dd}(x-y)\hat\psi(x)\hat\psi(y)\rangle\approx
\frac{1}{2}\int dxdy\,\langle\hat\psi^\dagger(x)\hat\psi(x)\rangle V_{\dd}(x-y)\langle\hat\psi^\dagger(y)\hat\psi(y)\rangle 
\end{equation}
Then, we discretize space into intervals of length $l$ indexed by $i$. We can rewrite Hamiltonian as a sum of a local term (short-range interactions and the kinetic energy of atoms) and a non-local term (dipolar interactions):
\begin{equation}
\langle\hat{H}\rangle=\sum_i\langle\hat{H}^i\rangle=\sum_i\langle\hat{H}_{\mrm {loc}}^i\rangle+\sum_{ij}\langle\hat{H}_\dd^{ij}\rangle\label{pomeq}
\end{equation}
Using our approximations we get:
\begin{equation}
\langle\hat{H}^i\rangle\approx\langle(\hat{T}^i+\hat{V}_\sr^i)\rangle+\frac{1}{2}\sum_j V_\dd((i-j)l)\langle\hat\psi^\dagger(j\cdot l)\hat\psi(j\cdot l)\rangle l.
\end{equation}
The last term of the above equation is constant within the $i$-th interval because we previously assumed that the dipolar potential and the density profile varies slowly.
Therefore, locally, the Hamiltonian admits the Lieb-Liniger form. The only departure from the Lieb-Liniger model is the lack of the periodic boundary conditions, which does not affect the spectrum significantly. In fact, the recent experimental study of the Lieb-Liniger spectrum was performed on a gas in a waveguide~\cite{Meinert2015}. 

All of the above provides sensible reasons to approximate the sum of the kinetic and short-range interaction energies in the $i$-th interval by the energy of the ground state of the Lieb-Liniger model~\cite{LiebLiniger1963}  calculated for the density $N|\psi(i\cdot l)|^2$:
\be
e_{\mrm {LL}}\left(|\psi|^2\right)=\frac{1}{2}N^2(N-1)|\psi|^6 e\left( \frac{g}{N|\psi|^2}\right) \label{liebs1},
\ee
where $e(\gamma)$  is a function defined in the Lieb-Liniger work~\cite{LiebLiniger1963}. Taking sum over all $i$'s we get:
\begin{equation}
\label{sum}
\langle\hat{H}\rangle\approx\sum_i e_{\mrm {LL}}\left(|\psi(i\cdot l)|^2\right)l+\frac{N(N-1)}{2}\sum_{ij}|\psi(i\cdot l)|^2V_{\mrm {dd}}((i-j)l)|\psi(j\cdot l)|^2l^2,\qquad \sum\limits_i|\psi(i\cdot l)|^2l\approx1
\end{equation}
which can be approximated by an integral
\begin{equation}
\begin{aligned}
\langle \hat H\rangle\approx&\int dx\, e_{\mrm {LL}}\left(|\psi(x)|^2\right)+\frac{N(N-1)}{2}\int dx\int dy\,|\psi(x)|^2V_{\mrm {dd}}((x-y))|\psi(y)|^2\\
&\int|\psi(x)|^2dx=1\label{finalfunctional}
\end{aligned}
\end{equation}
Note, that the last approximation \textbf{does not} correspond to $l\rightarrow0$; we only state that the value of sum \eqref{sum} is close to the value of the integral \eqref{finalfunctional} (which comes directly from assumptions \eqref{assume}).

Note also, that when making the approximation $|\psi(x)|^2=\rm{const.}$ in every interval $x\in[i\,l,(i+1)\,l]$, we have neglected the kinetic energy connected with changing its value between neighbouring intervals. To recompense this fact, we add the kinetic energy of the envelope of the state. We end with an energy functional:
\begin{equation}
\begin{aligned}
\label{last}
E\left(\psi(x)\right)=&\int dx\, \left(\frac{N}{2}\left|\nabla\psi(x)\right|^2+e_{\mrm {LL}}\left(|\psi(x)|^2\right)\right)+\frac{N(N-1)}{2}\int dx\int dy\,|\psi(x)|^2V_{\mrm {dd}}((x-y))|\psi(y)|^2\\
&\int|\psi(x)|^2dx=1
\end{aligned}
\end{equation}
The final equation resembles Gross-Pitaevskii energy functional (with additional term $e_{\mrm {LL}}\left(|\psi(x)|^2\right)$). It is also worth to note, that for small $g$ it reduces to this functional (as for $g\rightarrow 0$ in  local Lieb-Liniger model, the ground state tends to the product state where every particle is in the zero-momentum mode). In any regime of parameters, the functional \eqref{last} may be minimized by using analogous methods as for the Gross-Pitaevskii energy functional. Because of this mathematical similarity and to pay tribute to great physicists, we name corresponding differential equation (Eq. (4) in the main text) by the Lieb-Liniger Gross-Pitaevskii equation (LL-GPE).

We want to stress, that the only arbitrary assumption during the derivation was that the size of the bound state $d$ satisfies $d/N\ll \sigma$.
Therefore, for given interaction parameters, one has to minimize \eqref{last} with respect to the density amplitude $\psi(x)$ and check if the solution satisfies this assumption. If it does, the application of the equation \eqref{last} will be justified in that case. Otherwise, the set of parameters is outside of the range of applicability of our approximation and the results should be treated with very limited confidence.

\subsection*{Approximated formula for Lieb-Liniger energy}

To make derived functional useful for calculations, we need to have simple approximation for Lieb-Liniger energy, as there is no analytical formula for $e(\gamma)$. Therefore taking into account the following features:
\begin{itemize}
\item in the limit $\gamma\rightarrow 0$ it should agree with the first-order perturbation theory ($\frac{{\rm d}e}{{\rm d}\gamma}|_{\gamma=0}=1$)
\item in the limit $\gamma\rightarrow \infty$ it should reproduce Tonks-Girardeau energy $\lim_{\gamma\rightarrow\infty}e(\gamma)=\frac{\pi^2}{3}$
\item it should be smooth and convex in the intermediate region. 
\end{itemize}
 we roughly approximate it by

\be
e(\gamma)=\frac{\pi^2}{3}\frac{\gamma}{\frac{\pi^2}{3}+\gamma}.
\ee
With that, we rewrite Eq. \eqref{liebs1} into
\be
e_{\mrm {LL}}\left(|\psi|^2\right)=\frac{gN(N-1)}{2}\frac{|\psi|^6}{|\psi|^2+\frac{3g}{N\pi^2}}.\label{liebfunctional1}.
\ee

\section{Many-body Ansatz in the Tonks-Girardeau limit}
In the limit of infinitely strong interactions $g\rightarrow \infty$, we propose a variational Ansatz for the many-body ground state.
The Ansatz assumes, that the real ground state of our system, $N$ atoms with dipolar attraction and short-range repulsion is close to the ground state of the Tonks gas, but in the box of width d.
In such state one can compute the kinetic energy and interaction energy (short range and dipolar)  even for hundreds of atoms and perform minimization  with respect to the box width $d$. The details of the computation of the energies are presented step by step below.

Our system is characterized by:
\begin{itemize}
\item Open box  - hard walls at $x=0 $ and $x=d$
\item Inifnite contact interaction strength $g\rightarrow \infty$
\item Finite $g_\dd>0$
\end{itemize}
Our Ansatz assumes that the droplet state of such system is the ground state of infinitely repulsive bosons (fermionization) where the variational parameter $d$ depends on the dipolar interactions between the atoms. In the following subsections, we are going to introduce it in a more precise form. We use the box units as in the main text.

\subsection{Single particle eigenstates in the box with the open boundary conditions}
Single particle eigenstates and energies are given by:
\begin{equation}
\varphi_n(x) = \sqrt{\frac{2}{d}} \sin\bb{ \frac{\pi n}{d} x} \qquad E_n = \frac{\hbar^2 \pi^2  n^2}{2 m d^2},\qquad {\rm where}\qquad n = 1,\, 2, \ldots
\label{eq:varphin}
\end{equation}

\subsection{Ansatz for the two-particle system}

Tonks-Girardeau Ansatz for an eigenstate with the first particle in orbital $n$ and the second in  $m$:
\begin{equation}
	\psi(x, y) = | \varphi_n(x)\varphi_m(y) - \varphi_n(y) \varphi_m(x) |/\sqrt{2}
	\label{eq:GS-two-atoms}
\end{equation}
The interaction energy is given by:
\begin{equation}
E_{\rm int} = \int_{0}^{d} {\rm d} x \,\int_{0}^{d} {\rm d} y \,\bb{\psi(x, y)}^2  V_{\mrm {eff}}(x-y)
\end{equation}
For arbitrary interaction potential $V (x-y)$ (in our case $V_{\mrm {eff}}(x-y)$ from the main text), it is possible to simplify the expression by performing one integration. To do this, we will change variables:
\begin{eqnarray}
u & = & x-y\\
v & = & \frac{1}{2}\bb{x+y}
\end{eqnarray}
The new variables are chosen such that the surface element is unchanged, i.e. ${\rm d} x\,{\rm d} y =  {\rm d} u \,{\rm d} v $.
The inverse transform is simply
\begin{equation}
x  =  \frac{1}{2}\bb{2v + u}\qquad y  =  \frac{1}{2}\bb{2v - u }
\end{equation}
The interaction energy after the variable change will be 
\begin{equation}
E_{\rm int} = \int_{0}^{d} {\rm d} x \,\int_{0}^{d} {\rm d} y \,\bb{\psi(x, y)}^2  V_{\rm {eff}} (x-y) = \int {\rm d} u {\rm d} v \, \,\tilde{\psi}(u, v)^2  V_{\rm {eff}} (u),
\label{eq:Eint-uv-two-atoms}
\end{equation}
The integration region can be simplified using the symmetries $V_{\rm {eff}}(x-y)=V_{\rm {eff}}(y-x) = V_{\rm {eff}}(u)$ and $\psi(x, y) =  \psi(y, x)$. The latter implies, that the wavefunction in the variables $(u,v)$ is even with respect to $u$:
\begin{equation}
\tilde{\psi}(u, v) = \psi( v + \frac{1}{2}u, v - \frac{1}{2}u)  = \psi( v - \frac{1}{2}u, v +  \frac{1}{2}u)  = \tilde{\psi}(-u, v)
\label{eq:psi-uv-symmetry}
\end{equation}
Therefore the integration with respect to the whole diamond is twice the integration with respect to its part for $u>0$:
\begin{equation}
E_{\rm int}  = 2\int_{0}^d {\rm d} u \int_{u/2}^{d-u/2} {\rm d} v \, \,\tilde{\psi}(u, v)^2  V_{\rm {eff}}(u)
\label{eq:Eint-uv-two-atoms2}
\end{equation}
The result is
\begin{align}
E_{\rm int}&(m,n) = \frac{4}{d} \int_{0}^d {\rm d} u\,V_{\rm {eff}}(u)\, \label{eq:eint-mn} \frac{d-u}{d}\bb{1-\cos\bb{\frac{m\pi\, u}{d}} \cos\bb{\frac{n\pi\, u}{d}} }\\
& + \frac{4}{\pi\, d\, m\, n\, (m^2-n^2)} \int_{0}^d {\rm d} u\,V(u)\,\bb{\cos\bb{\frac{n\pi\, u}{d}} - \cos\bb{\frac{m\pi\, u}{d}}}\bb{m^3 \sin\bb{\frac{n\pi\, u}{d}} + n^3 \sin\bb{\frac{m\pi\, u}{d}}}
\end{align}

\subsection{Tonks-Girardeau ground state with $N$ particles}
Kinetic energy of $N$ atoms in the Tonks-Girardeau limit:
\begin{equation}
E_{\rm KIN} = N \sum_k P(k) \frac{\hbar^2 k^2}{2m} = N \sum_{n=1}^N \frac{1}{N} \frac{\hbar^2 \pi^2\,n^2}{2m\,d^2} = \frac{\pi^2 \hbar^2 N (N+1)(2N+1)}{12 \,m\,d^2}
\end{equation}
The total interaction energy is given by
\begin{equation}
E_{\rm INT} = \sum_{ij}\langle \hat{V}_{ij} \rangle,
\end{equation}
where $\hat{V}_{ij} = V_{\rm {eff}}(\hat{x}_i - \hat{x}_j)$ indicates the interaction potential between $i$th and $j$th particles. 
Due to indistinguishability the binary interaction energy is  the interaction energy between the "first" and the "second" atoms multiplied by the number of pairs:
\begin{equation}
E_{\rm INT} = \frac{N(N-1)}{2}\langle \hat{V}_{12} \rangle .
\label{eq:eint_12}
\end{equation}
The many body Ansatz for the ground state can be written as:
\begin{equation}
\psi_{\rm GS}(x_1,\,x_2,\,\ldots , x_N) = \frac{1}{\sqrt{N !}}\left|\mathcal{A} \left[ \phi_1(x_1) \phi_2(x_2) \ldots \phi_N(x_N)  \right] \right|,
\end{equation}
where the  $\mathcal{A}$ is antisymmetrization operator. Note, that the ground state for the box with open boundary condition is just the absolute value of the fermionic ground state. This relation does not hold for excited states. It is also not strictly  true for the ground state in the box with periodic boundary conditions. The interaction energy between the two particles is expressed by:
\begin{eqnarray}
\langle \hat{V}_{12} \rangle &=& \int {\rm d}x_1\ldots \int {\rm d}x_N V_{\rm {eff}}(x_1-x_2) |\psi_{\rm GS}(x_1,\,x_2,\,\ldots , x_N)|^2 = \\
&=&\frac{(N-2)!}{N!}\sum_{1\leq m < n\leq N}\int {\rm d}x_1\int {\rm d}x_2 V_{\rm {eff}}(x_1-x_2) \,\mathcal{A} \left[ \phi_n(x_1) \phi_m(x_2) \right]^2\\
&=& \frac{1}{N (N-1)}\sum_{1\leq m < n\leq N}\int {\rm d}x_1\int {\rm d}x_2 V_{\rm {eff}}(x_1-x_2)\, 2\,\bb{\frac{1}{\sqrt{2}}\mathcal{A} \left[ \phi_n(x_1) \phi_m(x_2) \right]}^2 \\
&=& \frac{2}{N (N-1)}\sum_{1\leq m < n\leq N}E(n,m)
\label{eq:eint-aux-mb}
\end{eqnarray}
where $E_{\rm int}(m,n)$ is the contribution to interaction energy coming from two particles, one on the $m$th energy level and one on $n$th energy level. The formula for $E_{\rm int}(m,n)$ is given in Eq. \eqref{eq:eint-mn}.
Combining Eq. \eqref{eq:eint-aux-mb} with \eqref{eq:eint_12} one gets:
\begin{equation}
E_{\rm INT} = \sum_{1\leq m < n\leq N} E_{\rm int}(m,n),
\end{equation}
Therefore, the total energy is given by:
\begin{equation}
E = \frac{\pi^2 \hbar^2 N (N+1)(2N+1)}{12 \,m\,d^2} + \sum_{1\leq m < n\leq N} E_{\rm int}(m,n)\label{eq:total-energy}
\end{equation}
Then, we numerically minimize Eq. \eqref{eq:total-energy} over the width $d$.

\section{Numerical diagonalization}

We access the ground state of the Hamiltonian (1) by diagonalization using the Lanczos algorithm. As we are interested in investigating bound states, boundary conditions should not have significant impact on the results and therefore, we assume the periodic boundary conditions for numerical convenience. We use the Fock space spanned by the plane-wave basis.
We choose the maximum single-particle momentum value which potentially may occur in calculations $2\pi k_{\rm max}$ and then we set the energy cutoff $E_{\rm cut}=(2\pi k_{\rm max})^2/2$, e.i. we consider only the Fock states with total the kinetic energy smaller or equal $E_{\rm cut}$.
Using the fact, that the total momentum of the system is conserved (due to the periodic boundary conditions), in calculating the ground state we only need to consider the states satisfying $\sum_{k=-k_{\rm max}}^{k_{\rm max}} 2\pi k N_k=0$, where $N_k$ is a number of particles in a given mode $k$.

We want to stress the fact, that to provide the same accuracy with an increasing number of particles, it is \textbf{not enough} to keep $k_{\rm max}$ constant -- one should rather ensure that the amount of available kinetic energy per particle is the same, which means $k_{\rm max}^2/N=\rm{const.}$

\begin{figure}[h!]
\begin{centering}
		\includegraphics[width=0.7\textwidth]{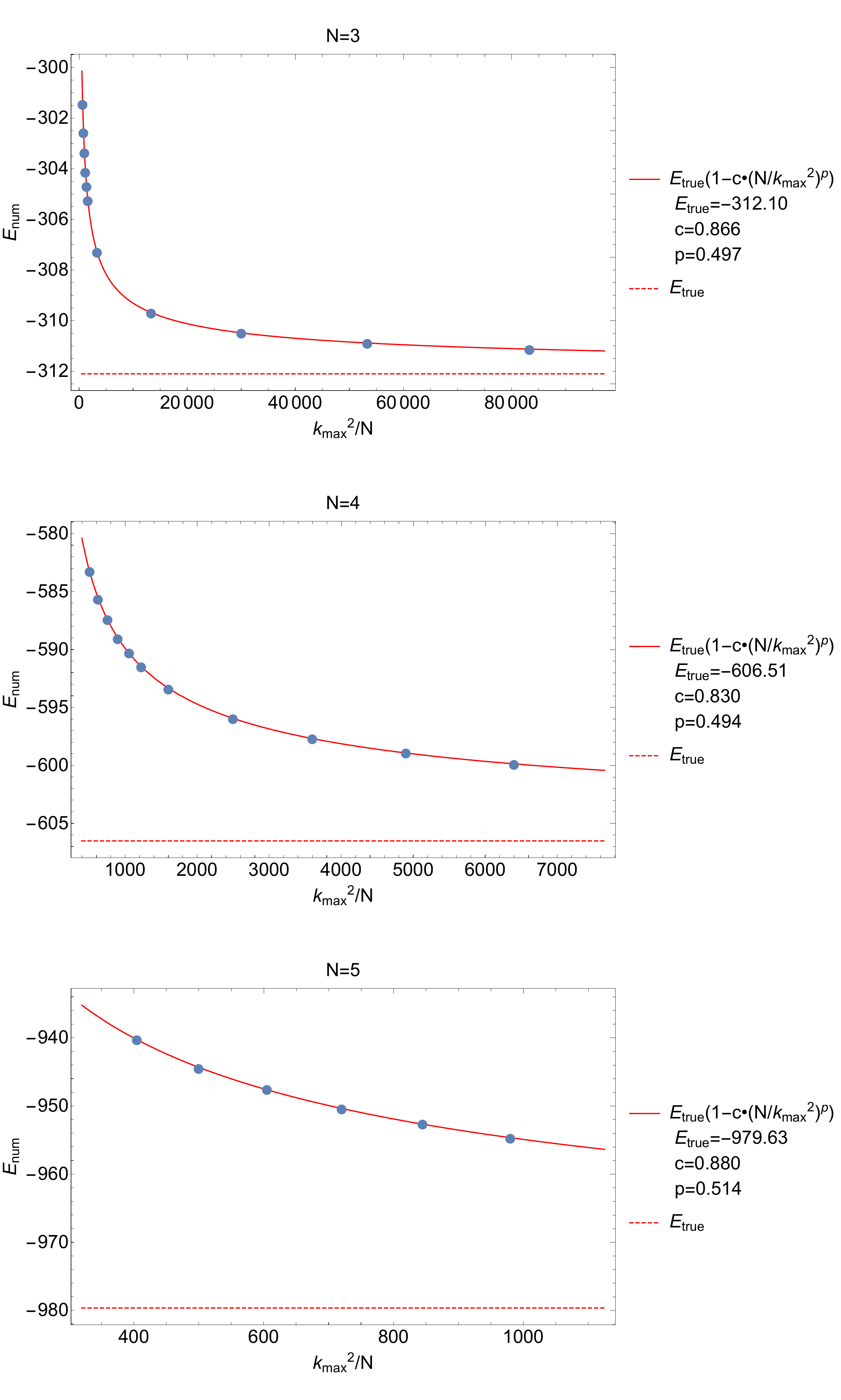}
		\end{centering}
		\caption{Above we present the energy convergence with increasing cutoff $k_{\rm max}^2/N$.\label{energy}}
\end{figure}

To get some information how far from the actual ground state we are,  we present in Fig. \ref{energy} the ground state energy as a function of increasing $k_{\rm max}^2/N$ for the droplet-like states from the main text (which was much more numerically challenging than soliton-like states). Assuming that a relative error decreases like some power of $k_{\rm max}^2/N$, e.i.
\begin{equation}
E_{\rm num}\left(\frac{k_{\rm max}^2}{N}\right)=E_{\rm true}\left(1-c\cdot \left(\frac{N}{k_{\rm max}^2}\right)^p\right)
\end{equation}
we are able to estimate the exact value of this energy. The fact that coefficients $c$ and $p$ estimated independently for $n=3,4,5$ does not varies significantly, supports reasonability of this assumption. From Fig. \ref{energy} we see that the energy of the ground state calculated with the cutoff $k_{\rm max}^2/N\approx1000$ is accurate up to $\approx 3\%$. For $N=5$ particles the number of Fock states needed for such accuracy is $4.8\cdot 10^5$. As a required dimension of Hilbert space increases exponentially with $N$, we are restricted to few particles calculations only.

\begin{figure}[h!]
\begin{centering}
		\includegraphics[width=0.6\textwidth]{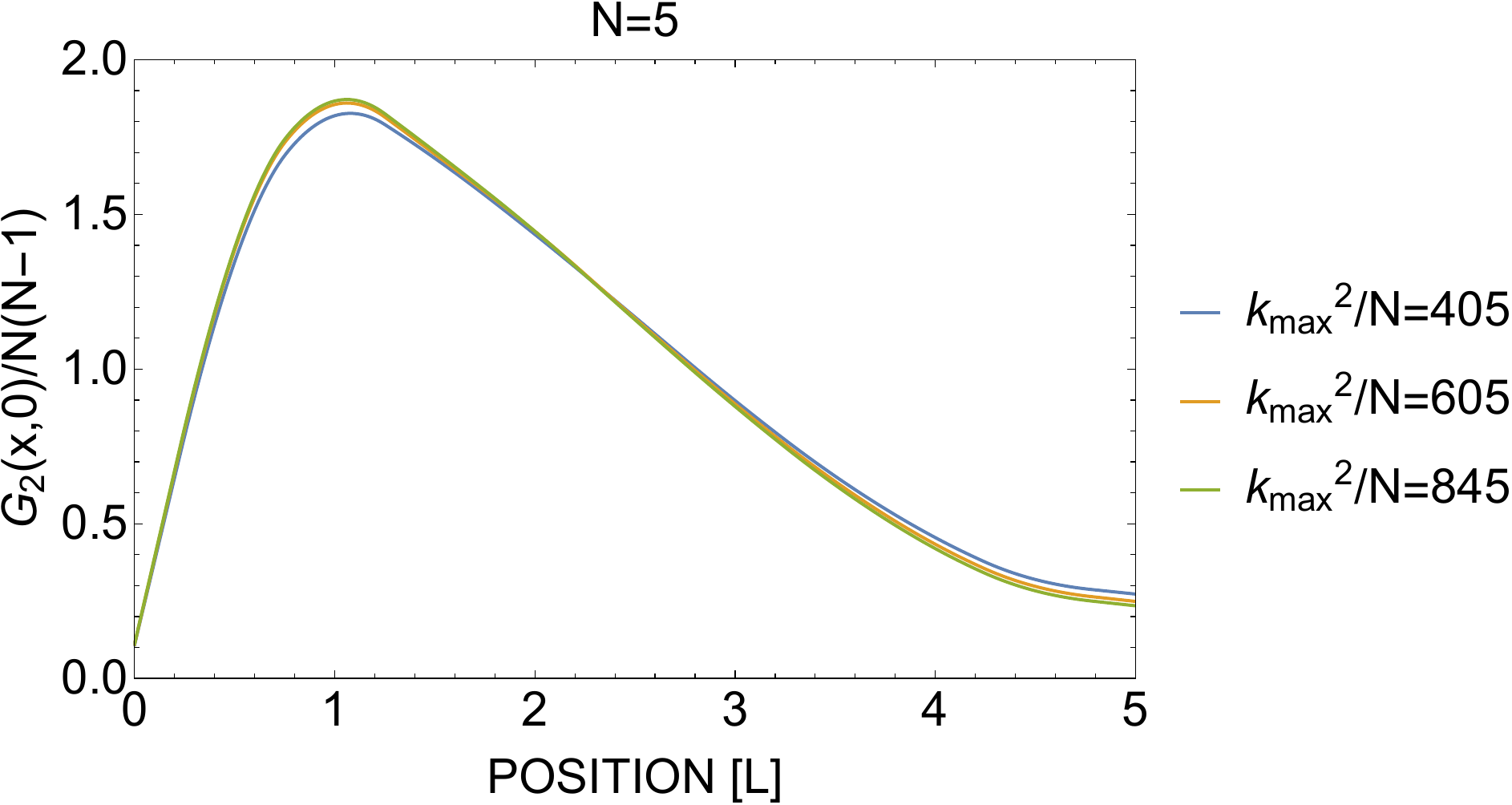}
		\end{centering}
		\caption{The changes of the shape of $G_2(x,0)$ with increasing cutoff are relatively small.\label{g2}}
\end{figure}

As we did not achieve full convergence for $N=5$, one may wonder if our discussion about this state (presented in FIG. 1 of main text) is reliable. To legitimate it, in FIG. \ref{g2} we show the changes of shape of second order correlation function $G_2(x,0)$ with the increasing cutoff. As these changes are relatively small, we find our results accurate enough to perform qualitative analysis presented in the main text.

\end{document}